\documentclass[conference]{IEEEtran}
\IEEEoverridecommandlockouts
\usepackage{cite}
\usepackage{amsmath,amssymb,amsfonts}
\usepackage{algorithmic}
\usepackage{graphicx}
\usepackage{textcomp}
\usepackage{xcolor}
\usepackage{newtxmath,mathtools}
\usepackage{algorithm}
\usepackage{multirow}
\usepackage{float}
\usepackage{caption}
\usepackage{subcaption}
\usepackage{array}
\usepackage{booktabs}
\usepackage{url}
\usepackage{makecell}

\usepackage{cancel}
\usepackage{ulem}
\def\BibTeX{{\rm B\kern-.05em{\sc i\kern-.025em b}\kern-.08em
    T\kern-.1667em\lower.7ex\hbox{E}\kern-.125emX}}
\begin{document}

\title{Using Deep Reinforcement Learning for mmWave Real-Time Scheduling}

\author{\IEEEauthorblockN{1\textsuperscript{st} Barak Gahtan}
\IEEEauthorblockA{\textit{Computer Science dept} \\
\textit{Technion}\\
Haifa, Israel \\
barakgahtan@cs.technion.ac.il}
\and
\IEEEauthorblockN{2\textsuperscript{nd} Reuven Cohen}
\IEEEauthorblockA{\textit{Computer Science dept} \\
\textit{Technion}\\
Haifa, Israel \\
rcohen@cs.technion.ac.il}
\and
\IEEEauthorblockN{3\textsuperscript{rd} Alex M. Bronstein}
\IEEEauthorblockA{\textit{Computer Science dept} \\
\textit{Technion}\\
Haifa, Israel \\
bron@cs.technion.ac.il}
\and
\IEEEauthorblockN{4\textsuperscript{th} Gil Kedar}
\IEEEauthorblockA{\textit{Ceragon Ltd.} \\
Israel \\
gilke@ceragon.com}
}

\maketitle

\begin{abstract} We study the problem of real-time scheduling in a multi-hop millimeter-wave (mmWave) mesh. We develop a model-free deep reinforcement learning algorithm called Adaptive Activator RL (AARL), which determines the subset of mmWave links that should be activated during each time slot and the power level for each link. The most important property of AARL is its ability to make scheduling decisions within the strict time slot constraints of typical 5G mmWave networks. AARL can handle a variety of network topologies, network loads, and interference models, it can also adapt to different workloads. We demonstrate the operation of AARL on several topologies: a small topology with 10 links, a moderately-sized mesh with 48 links, and a large topology with 96 links. We show that for each topology, we compare the throughput obtained by AARL to that of a benchmark algorithm called RPMA (Residual Profit Maximizer Algorithm). The most important advantage of AARL compared to RPMA is that it is much faster and can make the necessary scheduling decisions very rapidly during every time slot, while RPMA cannot. In addition, the quality of the scheduling decisions made by AARL outperforms those made by RPMA.   
\end{abstract}


\section{Introduction}
As mobile traffic increases, mobile networks must become more software-driven, virtualized, flexible, intelligent, and energy-efficient. 5G is the next-generation cellular network technology aiming to meet these requirements. 5G networks have flexible architectures and can support very high user density \cite{7390965}. To meet the high bandwidth demands, 5G networks use high-frequency mmWave communication ranging between 30 to 300 GHz. Because of the use of such high frequencies, mmWave signals are susceptible to shadowing and exhibit high attenuation; consequently, the scalability of 5G networks is presently a major challenge \cite{mezzavilla20155g,niu2015survey}.

Since mmWave signals require line of sight, 5G mmWave access networks in urban areas must use dense placement of base stations (BSs) compared to 3G and 4G networks. The need to connect a large number of BSs to the backbone imposes several significant technological challenges. The 3GPP standardization body proposes a multi-hop access architecture, called Integrated Access and Backhaul (IAB) \cite{atis3gpp-specification}, which aims to address these challenges, while imposing strict timing constraints on the network scheduler.

This paper shows that deep reinforcement learning (DRL) is a promising solution for addressing tough challenges related to network densification in a 5G multihop mmWave backbone, and in particular making scheduling decisions under strict timing constraints. The Open Radio Access Network (ORAN) alliance suggests conducting offline training using a realistic representation of the actual environment and deploying the learned policy to the open RAN Intelligent Controller (RIC)\cite{ORAN-WG2-AIML-V010202}. 

This paper considers a multihop mmWave transport wireless mesh, in which data packets are routed from the core network to the users and vice versa, via a multihop mmWave network. In this network, the BSs are used not only for connecting users to the network, but also as part of the multihop mmWave backbone \cite{TS-138-300-V16.2.0}. This paper proposes a model-free DRL algorithm \cite{degris2012model} for making fast scheduling and power assignment decisions, while conforming to the strict time constraints of a 5G mmWave network. To the best of our knowledge, this is the first paper to address these problems using DRL. 

The proposed DRL algorithm does not make assumptions about the interdependence of the various mmWave links, nor about the topology itself. It is trainable for a wide range of topologies and network interference models and is adaptable to a wide range of traffic workloads. It determines in real-time within the time duration of one slot, which BS pairs should activate their connecting mmWave links during each configurable time slot and using what power level.

\begin{figure}[tb]
    \centering
    \includegraphics[width=0.8\columnwidth]{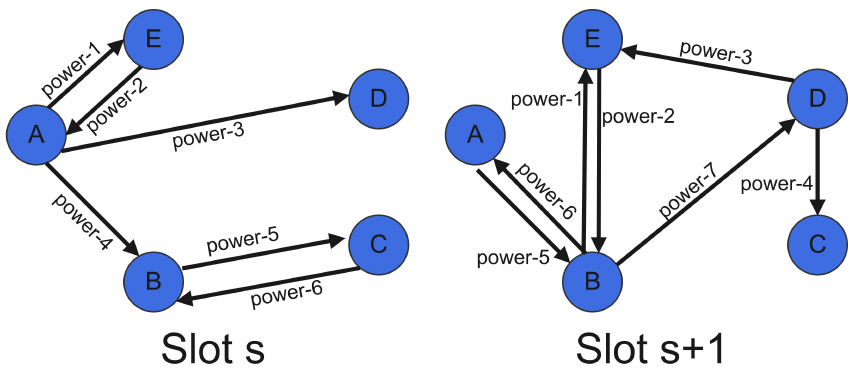}
    \caption{An illustration of the scheduling decisions made by the proposed algorithm during two consecutive time slots}
    \label{fig:slot}
\end{figure}

Figure \ref{fig:slot} illustrates a multihop mmWave backhaul with five BSs. It shows the mmWave links selected by the DRL algorithm to be active during two consecutive slots. In slot $s$, the algorithm activates the links from $A$ to $B,D$ and $E$; the link from $E$ to $A$; the link from $B$ to $C$; and the link from $C$ to $B$, each with its own power. In slot $s+1$, the DRL algorithm activates a different set of links, with different transmission powers.

The rest of this paper is organized as follows. Section \ref{RELATEDSEC} discusses the related work. Section \ref{MODEL} introduces the considered network and interference models. Section \ref{SOLUTION} presents the AARL algorithm. Section \ref{GREEDYSEC} outlines a benchmark algorithm whose performance is compared in Section \ref{EVALSEC}, to that of AARL. Finally, Section \ref{CONCLUDESSEC} concludes the paper.


\section{Related Work}
\label{sect:related work}
While many papers have examined how to use machine learning for routing or scheduling in communication networks \cite{valadarsky2017learning,valadarsky2017machine,xie2018routenet} only a few papers have explicitly addressed the problem of routing or scheduling in a multihop mmWave backbone while taking into account non-trivial interferences \cite{8377239,wang2018deep}.

In \cite{9653135,hasanzadezonuzy2020reinforcement}, DRL and gradient decent are used to maximize the throughput of unicast flows in multihop mmWave networks. The authors consider strict per packet deadlines. Their model is different from ours in several aspects: Firstly, they do not consider the interference between links and therefore do not address the problem of power control. Secondly, they assume that a single slot is sufficient for delivering every packet to its final destination, while our paper abandons this unrealistic assumption.

In \cite{9468707}, DRL is used for channel and latency-aware radio resource allocation. This work aims to optimize the uplink scheduling for service-oriented multi-user mmWave RAN in 5G. Multiple application flows are implemented with various statistical models, and the key function modules of the system are designed to reflect the operation and requirements of service-oriented RANs. The characteristics of the mmWave channels utilized in the considered system are collected experimentally and verified via a radio-over-fiber mmWave testbed with dynamic channel variations. In contrast to \cite{9468707}, our work focuses on scheduling in the mmWave backhaul, which is a different problem with different requirements.

In \cite{8757174}, deep deterministic policy gradient (DDPG) is used for radio resource scheduling in a 5G RAN. This work considers multiple users, variable channel conditions, and random traffic arrivals. The limitation of this work is that only Poisson distribution is employed to model the arrival of users. Unlike our work, \cite{8757174} does not address the backhaul scheduling problem.

In \cite{feng2019dealing}, DRL is used for addressing the problem of random blockage of mmWave access links and its effect on the backhaul. The authors learn the blockage pattern and show that the system's dynamics can be captured and predicted, resulting in a better utilization of backhaul resources. While both \cite{feng2019dealing} and our paper use DRL for solving 5G backhaul-related problems, the specific problems addressed by the two papers are substantially different.

In \cite{rischke2020qr}, an algorithm called QR-SDN is introduced. QR-SDN is a tabular RL algorithm, which directly represents the routing paths of individual flows in their state-action space. Unlike the present study, the authors do not consider the interference between the links. Moreover, their algorithm may move a flow from one path to another, while our algorithm sticks to the same path to prevent out-of-order routing.

In \cite{xu2018experience}, a highly effective DRL-based model-free control framework is developed for traffic engineering in a general (not 5G) communication network. The authors consider a general communication network and show how DRL can enable experience-driven networking. The proposed algorithm can be used for learning network dynamics, and for making routing decisions. They use DRL as a model-free routing method, while ignoring interferences.

The present study can also be viewed as an approach to addressing a resource allocation problem using DRL. In this context, \cite{he2017deep} previously used DRL for estimating the availability of cache and selecting a proper set of users for interference alignment. In  \cite{wang2018deep}, DRL was used for solving the problem of multi-channel access, where every user observes the history of the channel's dynamics and predicts the possible actions of other users. In \cite{8969579}, the authors used DRL for bridging the gap between different performance requirements in a datacenter network. They considered multiple resources, such as bandwidth, cache, and computing.\label{RELATEDSEC}

\section{Network Model and Problem Statement}\label{MODEL}
\begin{figure}[tb]
    \centering
    \includegraphics[width=0.6\columnwidth]{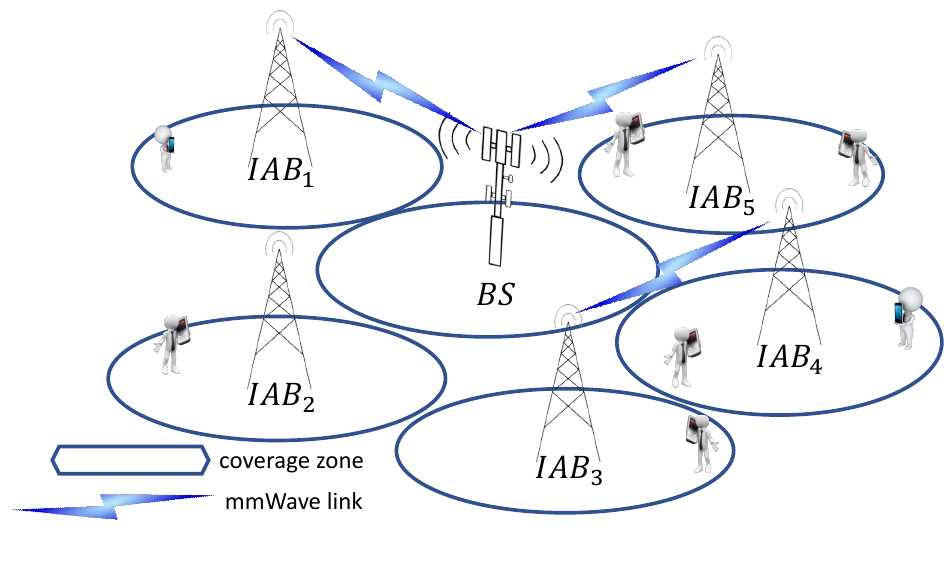}
    \caption{A multihop mmWave wireless backhaul with five IAB nodes and one BS}
    \label{fig:map1}
\end{figure}

\textbf{The Network Model}
Following \cite{7925837,lukowa2020coverage,8422149,saha2019millimeter, zhang2020rl}, we consider a multihop mmWave mesh, where data packets are received from the core network and are routed to and from the users via one or more 5G mmWave links. In 5G, two types of mmWave nodes exist, BSs and IABs. The difference between the two is that an IAB is connected to other mmWave nodes, and to some users, while a BS is connected only to other mmWave nodes (see Figure \ref{fig:map1}). Since this paper focuses on the transition of packets between mmWave nodes, and not between the users and these nodes, we do not distinguish between an IAB and a BS, and refer to both as BSs.
\label{sec:length}

In the considered network model, the RIC receives a demand vector $v_i$ from every $BS$ before the beginning of every time slot. The RIC determines which mmWave links will be established during the next slot, and using what power. Each $BS_i$ has a buffer of waiting packets associated with every neighbor $BS_j$. This buffer is referred to as $B_{i \rightarrow j}$. If an mmWave link is established from $BS_i$ to $BS_j$ during some slot, this link is used for transmitting packets from $B_{i \rightarrow j}$ to $BS_j$. The demand vector $v_i$ sent by $BS_i$ to the RIC indicates the number of packets that $BS_i$ needs to transfer to each neighbor $BS_j$. 

Since sending the demand vector from each BS to the RIC takes some non-negligible time, each BS is required to send its demand vectors for slot $s+1$ at the beginning of slot $s$. For every neighbor $BS_j$, this vector indicates the number of packets waiting to be transmitted from $BS_i$ to $BS_j$ at the beginning of slot $s$. Since the RIC knows which links are activated during slot $s$, it can estimate the number of packets that $BS_i$ will be able to transmit to every neighbor during slot $s$, and therefore the number of waiting packets remaining in the buffer of $BS_i$ will have just before slot $s+1$ starts. The decision which links to activate during slot $s+1$ is made by the RIC after it gathers the demand vectors from all BSs. The demand vectors and the controller's scheduling decisions are exchanged over a special control channel, such as 4G-LTE channels \cite{giordani2016uplink, gures2020comprehensive}, independant of the mmWave infrastructure.

Routing is always made on the shortest paths for the following reasons: (1) to ensure in-order delivery of packets belonging to the same flow; (2) to minimize bandwidth consumption and maximize total throughput; (3) to prevent routing loops without using a complex routing protocol; and (4) to reduce the dimension of the DRL optimization problem. 

Suppose that the RIC decides to activate the link from $BS_i$ to $BS_j$ during a specific slot $s$. This implies that during this slot, packets are transmitted from $BS_i$ to $BS_j$. These packets are received by $BS_i$ either before or during slot $s$. We do not consider a layer-2 flow control protocol due to the negative impact of head-of-the-line blocking on the total throughput. Thus, it is possible that some of the packets sent by $BS_i$ to $BS_j$ do not have an available buffer capacity and are therefore dropped by $BS_j$.

The most important component of an ML scheduling algorithm is the definition of an appropriate optimization criterion. In our case, the minimization of the average delivery time is not a good choice, as it encourages the algorithm to drop many packets to minimize the delivery time of the delivered packets. Maximizing the total throughput is a better option, but the total throughput depends on the total load, which, in turn, depends on the congestion control protocol. With this protocol, the load increases when the packet loss decreases and vice versa. Thus, we believe that the best criterion to minimize is the total number of dropped packets.

\begin{figure}[tb]
    \centering
    \includegraphics[width=0.6\columnwidth]{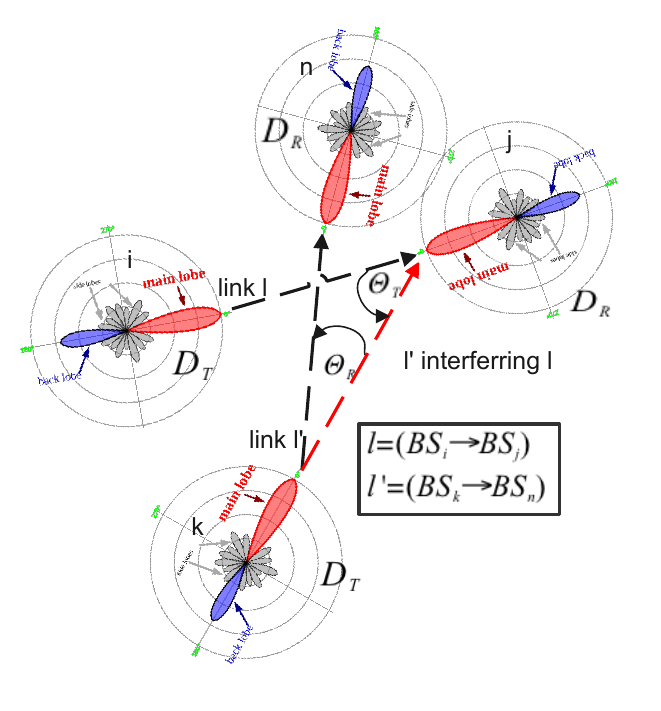}
    \caption{The interference between two active mmWave links}
    \label{fig:map555}
\end{figure}

\textbf{The Interference Model}
Let $l$ be the mmWave link from $BS_i$ to $BS_j$. Let the transmission power of $BS_i$ be $P_{i}(l)$. The received power at $BS_j$ can be calculated according to the Friis transmission equation \cite{shaw2013radiometry}. The equation defines the free space path loss (FSPL) as a function of the transmit power $P_{i}(l)$, the received power at $BS_j$, the distance between the two antennas, and the antennas' gains in a free-space communication link: $\mathrm{FSPL(i,j)} = D_\mathrm{T}D_\mathrm{R}\left(\frac{c}{4 \pi  D(i,j) f}\right)^{2}.$ Here, \textit{c} denotes the speed of light, $D(i,j)$ is the distance between $BS_i$ and $BS_j$, $D_\mathrm{T}$ and $D_\mathrm{R}$ are the directivities of the transmitting and receiving parabolic antennas respectively, and $f$ is the radio carrier frequency. FSPL represents the loss factor depending on the transmit distance and wavelength. According to the radiation pattern function, the received signal $D_\mathrm{R}$ is affected by the signal angle of arrival, where $\theta_\mathrm{R}$ is the deviation of the beam direction from the normal to the receiver plane, in the following way:  $D_\mathrm{R} \propto e^{-\frac{4(\theta_\mathrm{R})^{2}}{\sqrt{2}}}.$ The received power at $BS_j$ is given by, $P_\mathrm{R_{j}}(l) = P_{i}(l) - \mathrm{FSPL(\textit{i,j})}-\eta$,
where $\eta$ is a uniform thermal noise. $D_\mathrm{T}$ is defined similarly using the transmit angle $\theta_\mathrm{T}$ (refer to Figure \ref{fig:map555}).

Let $l'$ be another link transmitting from $BS_k$ to $BS_n$ at power $P(l')$. This link induces an interference in link $l$, which is defined as $\mathrm{I}(l,l')$. Assuming that all interferences are summed up adversarially, we define the effective receive power of link $l$ as
\begin{equation}
   P_\mathrm{R_\mathrm{eff}}(l) =  P_\mathrm{R}(l)  - \sum_{l' \ne l} \mathrm{I}(l,l').
   \label{eq:effective}
\end{equation}
Thus, the actual capacity of link $l$ is given by $C(l) = B \cdot log_{2}(1+P_\mathrm{R_\mathrm{eff}}(l))$,
where $B$ is the link's nominal bandwidth  (in the absence of any interference). The number of bits that can be transmitted over $l$ is given by $N_l = C_l \Delta t$,
where $\Delta t$ is the configurable slot duration. Figure \ref{fig:map555} illustrates the interference between links $l$ and link $l'$. When network interference level reaches 100\%, only one link can transmit at a time.

\textbf{Integration into ORAN deployment}
The ORAN Alliance proposes an architectural innovation based on two core principles \cite{bonati2021intelligence}. Firstly, it promotes a model where the BSs are virtualized and divided across multiple network nodes. Secondly, it defines the RAN-RIC, which provides a centralized abstraction of the network, and allows operators to implement and deploy custom control plane functions. 

ORAN's architecture defines three control levels for different timing constraints: real-time (RT), near-real-time (near-RT), and non-real-time (non-RT). RT operations are carried out by the network elements, whereas non-RT and near-RT operations are carried out by the RIC. While the training of an ML algorithm is conducted at the non-RT level, the inference of trained neural network is deployed at the near-RT level, where it interacts with the network elements, receives their measurements, and controls their configurations and policies \cite{polese2022understanding}.

\begin{figure}[t]
    \centering
    \includegraphics[width=0.5\columnwidth]{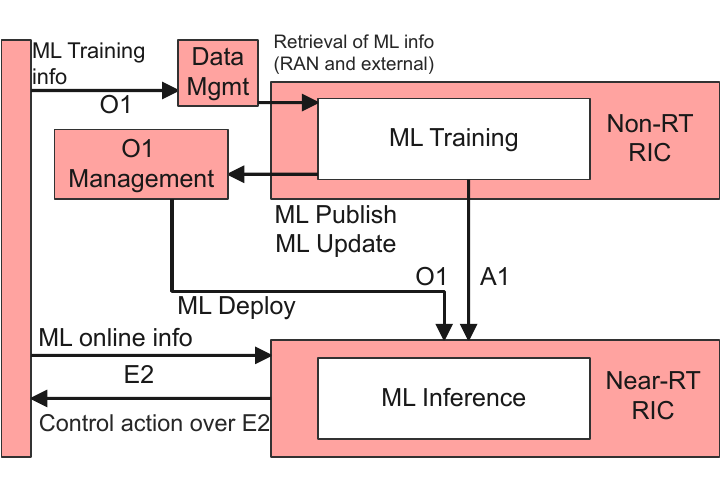}
    \caption{ORAN deployment scenario \cite{ORAN-WG2-AIML-V010202}}
    \label{fig:oran-ifer}
\end{figure}

Figure \ref{fig:oran-ifer} illustrates ORAN's training phase and an inference phase. Specifically, the considered deployment of the proposed algorithm corresponds to scenario 1.2 in Section 4.1 of \cite{alliance2019ran}. During the training phase, the environment provides AARL with the demand vector of the BSs. During the inference phase, the weights of the AARL's neural network are published to the near-RT RIC via the O1 interface. AARL operates at the near-RT, where it receives the vector demands from all BSs via the E2 interface prior to the beginning of the next time slot and sends the scheduling decision to the BSs via the E2 interface. As stated in Chapter 5 of \cite{alliance2019ran}, the near-RT loop must adhere to the time of 10ms time budget.

\section{AARL Scheduling Algorithm}\label{SOLUTION}
Scheduling is a decision-making activity, which is typically represented and solved as a utility maximization problem. The formulation of the problem is based on a mathematical model, which requires assumptions and approximations. In a 5G scenario, numerous aspects, including the users' mobility pattern and the traffic model influence the system modeling. Moreover, since data rates are very high, it is critical to meet the strict timing constraint for each radio time slot. These aspects make the multihop mmWave communication network a highly dynamic system. Hence, it is nearly impossible for such a model to account for all these and other factors. 

DRL offers an innovative approach to solving decision-making problems in the absence of an explicit mathematical model. A DRL agent approaches the optimal solution of a Markovian decision process by learning from its interaction with its environment, and is trained to make decisions that maximize the reward function. Once trained, the agent can make very fast scheduling decisions, which are also very effective.

Existing solutions for scheduling usually rely on network information exchange, resulting in a trade-off between overhead and performance. In contrast, DRL seeks to optimize the network tasks through a trial-and-error process that does not require explicit or instantaneous network information. A DRL algorithm can be model-free, which means that it does not require explicit knowledge of network topology or the interdependencies between different mmWave links. Therefore, we believe that a DRL-based algorithm is an excellent choice for performing very fast and effective scheduling and power control decisions in a multihop mmWave wireless network.

To use AARL, an operator needs to load the BS locations and interference model into the simulated environment and train it offline. Once the agent is trained, the trained neural network should be loaded into the RIC controller.

The environment and agent definitions used by AARL for deciding which mmWave links will be activated during each slot are as follows: 
\begin{itemize}
  \item Training is divided into episodes. Each episode is a sequence of slots required to completely deliver a traffic matrix. A traffic matrix indicates the amount of packets with a source and destination. When an episode starts, the agent is given a traffic matrix and an interference matrix. Each entry in the interference matrix $I_{i,j}$ indicates the amount of interference mmWave link \textit{i} induces on mmWave link \textit{j}. Packets that are dropped or successfully delivered are removed from the system. The interference matrix is assumed to be fixed for the entire episode because of episode's relatively short duration. 
  \item The episode ends when each packet either reaches its destination or is dropped. 
  \item Each slot is a system step. After each step, the agent receives a reward for making good scheduling decisions during that step. 
  \item The agent knows which link each packet should traverse to reach its destination over the shortest path.
  \item Establishing the interference model and removing or dropping packets from various buffers are executed by the DRL's environment. The environment validates every decision the agent makes.
  \item During each step, multiple packets can move between neighboring BSs over activated links. However, each packet can move only one hop per step. This limitation is relevant only to the DRL. During a real-time slot, a packet can be routed multiple times, if the relevant links are activated. 
  \item The maximum capacity of each link is determined by the given topology. Throughout the episode, the interference model affects each step dynamically, by reducing the capacity of each activated link according to its power and the power of the other activated links.
 \end{itemize}
When using an on-policy DRL algorithm such as PPO, the agent learns from consecutive steps by assessing the quality of its decisions using a reward function. Thus, the best policy is learned by taking a small step at a time, while improving the current agent's policy. The goal of the agent is to find the best strategy suited for the environment. AARL employs a PPO algorithm for finding a good schedule for each step under varying interference and workload distributions. 

We trained AARL on a variety of traffic and interference matrices to ensure that it will make good scheduling decisions also for workloads unseen at training. The training set comprised a uniform workload of packets with varying interference levels. However, as we demonstrate in the evaluation phase, the method performs well also on highly skewed workloads, e.g., when all the packets originate from 10\% of the BSs and are destined to 90\% of the BSs, or vice versa. 

\begin{figure}[t]
\centering
\captionsetup{justification=centering}
\begin{subfigure}{0.40\columnwidth}
  \centering
  \captionsetup{justification=centering}
  \includegraphics[width=0.8\textwidth]{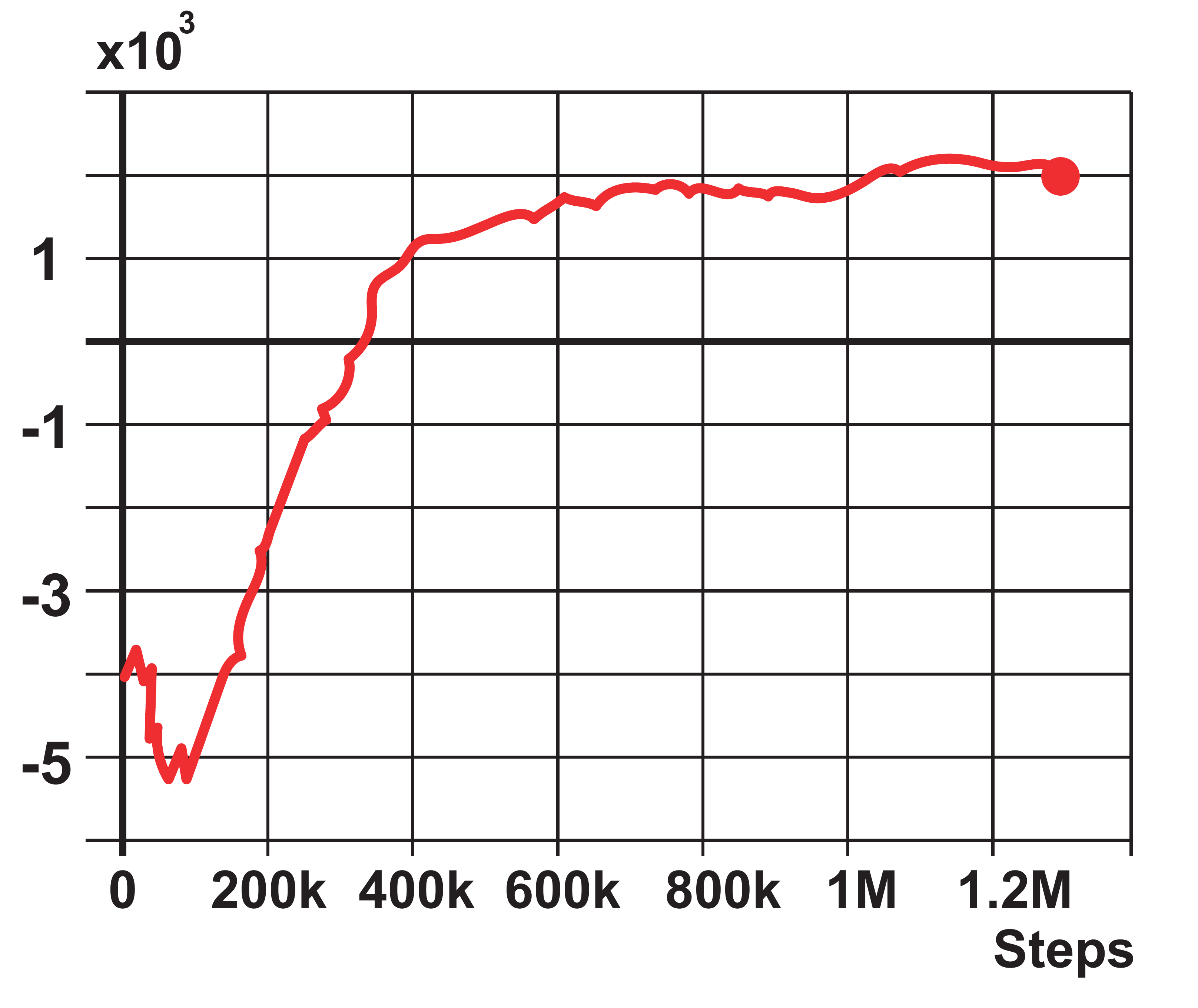}
  \caption{A convergence of the reward function}
  \label{fig:reward-convergence}
\end{subfigure}%
\begin{subfigure}{0.60\columnwidth}
  \centering
  \captionsetup{justification=centering}
  \includegraphics[width=1\textwidth]{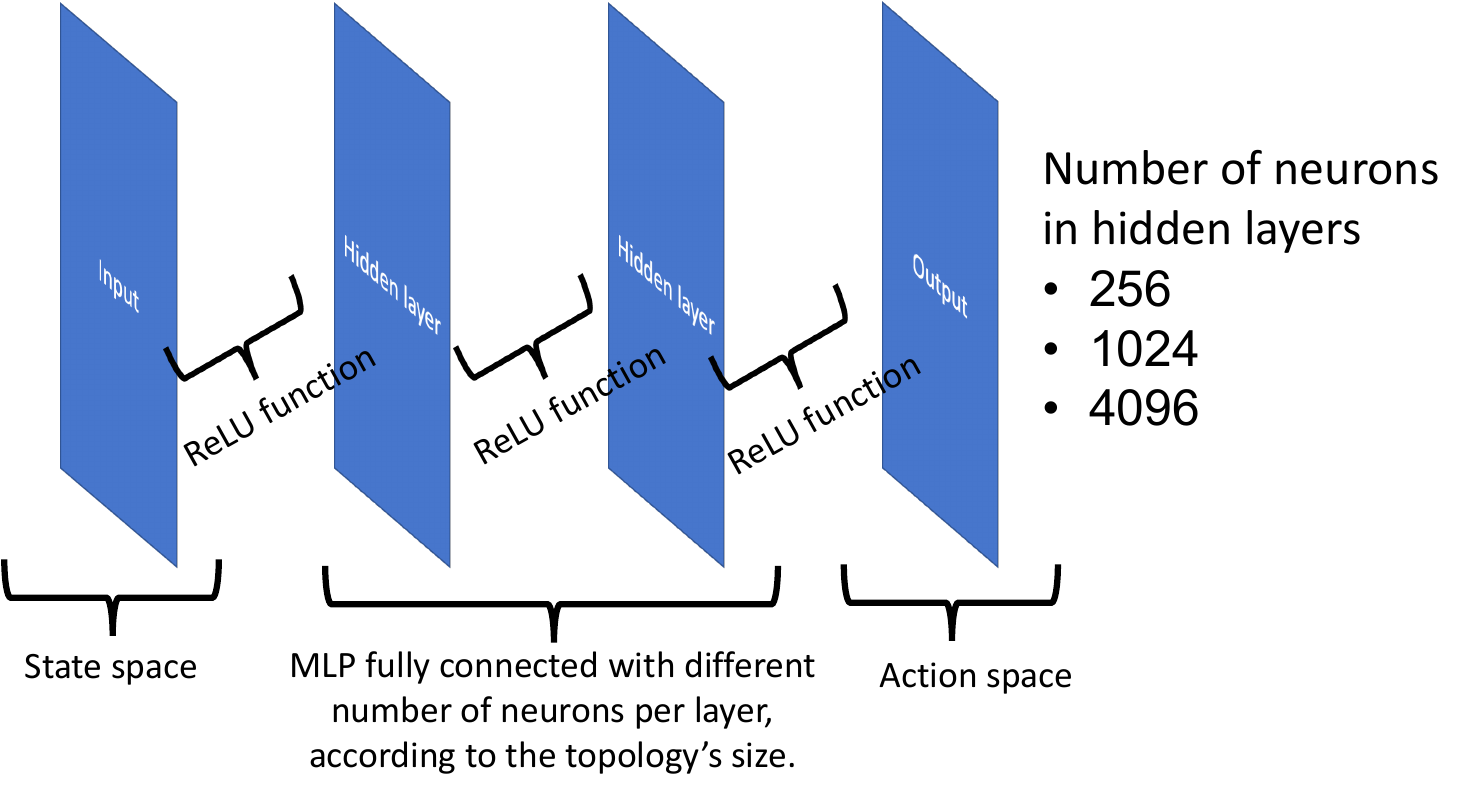}
  \caption{The neural network used by AARL}
  \label{fig:RL-NNN}
\end{subfigure}
\caption{Illustration of the DRL's formulation}
\label{fig:formulation-reward}
\end{figure}

Training was done on uniform workloads for the following reasons: 
\begin{enumerate}
    \item Exploration -- It allows the algorithm to explore a wider range of states and actions, which can help to prevent it from getting stuck in a local optimum, and increase the likelihood of finding a good global optimum;
    \item Generalization -- A uniform distribution of workload allows the algorithm to learn a more general policy, which can be applied to a broad range of situations.
    \item Robustness -- A DRL algorithm that has been exposed to a diverse set of states and actions is less likely to overfit to a specific set of conditions, making it more robust and less sensitive to small changes in the environment.
\end{enumerate}
Training is conducted over a period of several consecutive episodes. AARL converges the following way: First, AARL converges to a steady reward as shown in Figure \ref{fig:reward-convergence}. Each training run of an agent is sampled every 50,000 steps. After sampling, the best checkpoint of weights is taken by looking at the reward's function graph. The higher the plot, while being stable and not over-fitting, the better the checkpoint is.

The state space of the DRL is concatenated from several vectors. The first part of the state vector indicates the current load on each buffer. The \textit{i}'-th entry of this vector represents the load fraction for the \textit{i}'-th buffer of a specific mmWave link. The loads ranges between $[0,1]$, where $0$ indicates an empty buffer and $1$ indicates a full one. The second component of the state vector indicates the percentage of packets waiting in each buffer in proportion to all the packets in the system. For instance, if there are $1,000$ packets in all the buffers and only 10 in the considered buffer, this entry for the considered buffer is $0.01$. The third state vector component is the interference matrix, which is fixed for the entire episode.

The action space contains a vector indicating the power assigned to each mmWave link. Each entry in the action vector corresponds to a specific mmWave link in the topology. Here, 0 indicates that the link is not chosen for this step, and 1 indicates that it is chosen to work at maximum power. RPMA algorithm (described in Section \ref{GREEDYSEC}) uses power steps of 0.1 between 0 and 1, while AARL uses 0.01 steps. Hence, the total number of power levels is 11 for RPMA and 101 for AARL. This is because the run time of AARL is not affected by the number of power levels, whereas that of RPMA is dominated by this number.

As shown by Figure \ref{fig:RL-NNN}, the state space is AARL's input, and it is composed of several load vectors from the system. Depending on the size of the topology, the middle layers are Multi-Layer Perceptron (MLP) fully connected neural networks with 256 to 4,096 neurons connected by a ReLU activation function. The output is the action space, which is the selected set of mmWave links and their selected power level. The use of a small MLP neural network makes AARL very efficient when scheduling decisions must be taken very rapidly.

Developing a good reward function while promoting the minimization of the number of dropped packets is a major obstacle of an efficient DRL-based solution. Assuming that most connections are regulated by a TCP (or TCP-like) congestion control, minimizing the number of lost packets will maximize the total throughput. Our reward function for each step \textit{t} is therefore: $R_{t} = -\beta - \alpha \frac{D}{P} + \frac{M}{P},$
where \textit{P} is the total number of packets in the system before the step starts (i.e., the traffic load), and \textit{D} is the total number of packets dropped during the step, assuming that a packet is dropped if it is forwarded to the next hop but cannot be stored there due to buffer overflow. Thus, $\frac{D}{P}$ is the fraction of dropped packets. The reward function should be inversely proportional to it. 

We added another component to the reward function to encourage AARL to move as many packets as possible during each step. The total number of packets in-transit during the step is denoted by \textit{M}. Thus, $\frac{M}{P}$ represents the fraction of moving packets, and the reward is proportional to it. $\alpha$ is a scaling factor that indicates the importance of the first term over the second. Finally, the $\beta$ term aims to penalize the system for each additional step required to deliver the traffic demand.  We trained the system with two $\alpha$ values: $1$ and $10$, and one $\beta$ value of $1$.

In addition to the chosen reward function, several other reward functions were tested. Their results are not reported in this paper due to their insufficient performance. A reward function that demonstrated performance comparable to the selection function was $r_t = -p - s_{t}$. Here, \textit{p} is the penalizing term as explained before, scaled with three different values \{-1, -0.1, -10\} for each additional step required to deliver the traffic matrix. $s_{t}$ is the number of steps each packet was delayed relative to the shortest path it should have taken without delay. 

Recall that to use AARL, an operator needs to load the BS locations and interference model into the simulated environment and train it offline. The trained neural network should be loaded into the RIC controller. We trained AARL on three topologies provided by Ceragon Ltd. (see Figure \ref{fig:topologies}):

\begin{figure}[t]
    \centering
    \includegraphics[width=0.7\columnwidth]{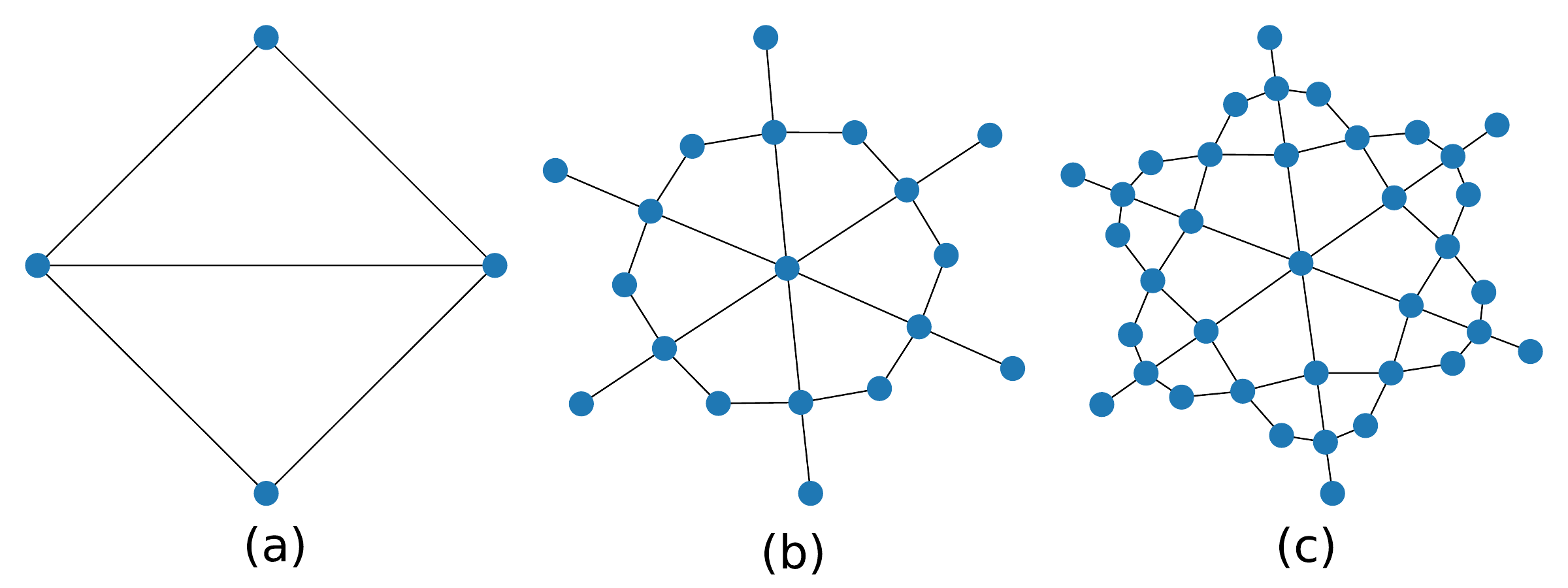}
    \caption{The three topologies used for AARL's training}
    \label{fig:topologies}
\end{figure}
\begin{description}
    \item[(a)] A topology with 4 nodes and 10 links. 
    \item[(b)] A topology with 19 nodes and 48 links, which represents a medium-size ORAN backhaul network. 
    \item[(c)] A topology with 37 nodes and 96 links, which represents a large-size ORAN backhaul network.
\end{description}
\begin{figure}[tb]
    \centering
    \includegraphics[width=0.8\linewidth]{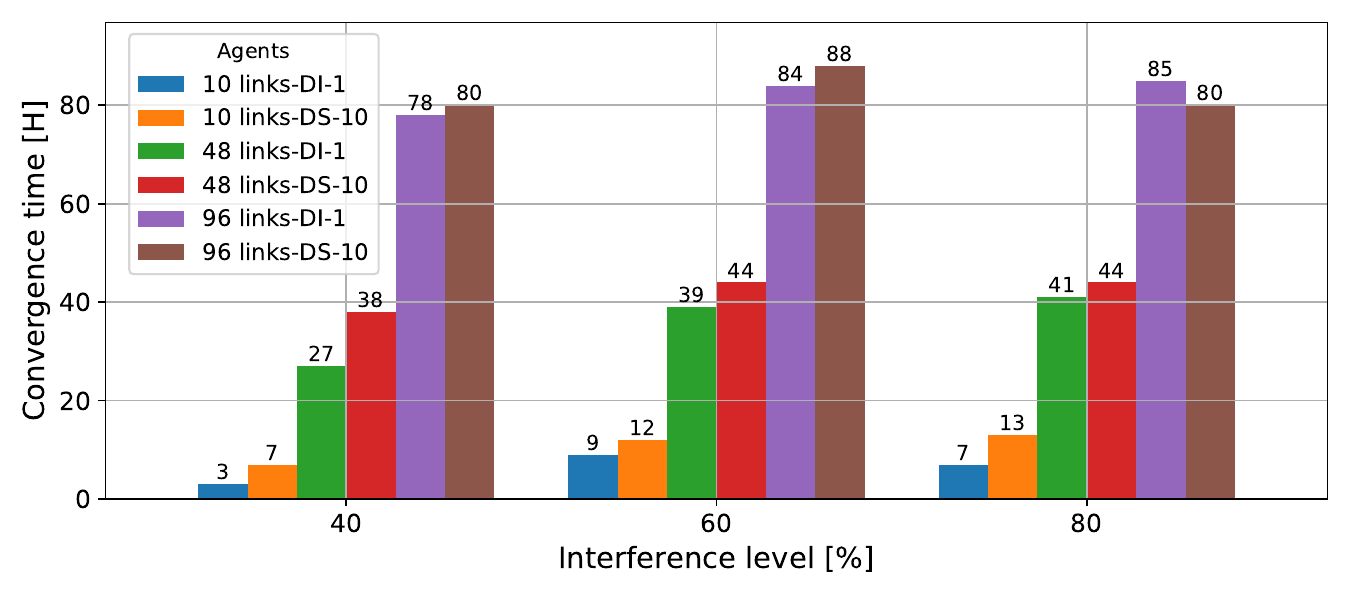}
    \caption{The convergence time of the various AARL algorithms for different interference levels, with a uniform workload}
    \label{fig:training-Convergence}
\end{figure}

A unique interference matrix was generated for each episode using the interference model described in Section \ref{MODEL}. AARL was trained with data packets that were generated from each BS uniformly. When training starts, the capacity of each mmWave link is sampled from a uniform distribution of $[115,125]$ packets per time step. Each buffer in the system has a capacity of 650 packets, the size of the buffer can be configurable. AARL is trained using Stable Baselines3 PPO agents \cite{Raffin_Stable_Baselines3_2020} in the following way; Five different PPO agents are trained, each with one of the following network interference levels: 20\%, 40\%, 60\%, 80\%, and 100\%. Recall that when the interference level reaches 100\%, only one mmWave link can transmit at full power. During training, the learning rate was set to $3\cdot10^{-4}$.

Figure \ref{fig:training-Convergence} shows the time to convergence of the training phase of the various AARL algorithms for different interference levels (see Figure \ref{fig:reward-convergence} for convergence). For example, for a topology with 96 mmWave links, AARL with $\alpha$=1 requires 85 training hours at  interference level of 80\%. For a topology with 48 mmWave links, AARL with $\alpha$=1 requires 41 hours at the same interference level.

The training time appears to be dominated by the topology size and the network load. As the topology size increases, the state and action spaces increase as well. For example, the action space is a vector of 96-dimensions for the 96-link topology, since it has an entry per each mmWave link. Similarly, for the 48-link topology, the vector has 48 dimensions. The implication on the state space is as follows. For the 48-link topology, the state space has 3 vectors: the first is with 48 entries; the second is also with 48 entries; and the third is with $48\cdot48=2,304$ dimensions. Therefore, the total dimension of the state space is $48+48+(48\cdot48)=2,400$. In contrast, the 96-link topology has two vectors each with a size of 96 entries, and a vector of $96\cdot96=9,216$ entries. Therefore, the state space dimension is $96+96+(96\cdot96)=9,408$. When the spaces are larger, the agent needs to learn more complex policies, requiring longer offline training time.

Additionally, as the level of interference increases, an episode length increases, since fewer mmWave links can transmit during each step. This phenomenon is illustrated by Figure \ref{fig:training-Convergence}. For example, the 10-AARL-DS agent converges after 7 hours for the 40\% interference level, after 12 hours for the 60\% interference level, and after 13 hours for the for the 80\% interference level. The trend is similar for all other agents. The length of an episode is another statistic that helps to decide when to stop training. When looking at the collected statistics from  RPMA's runs, we compared the number of steps before completion. The smaller this number, the better the network was utilized, as long as the number of dropped packets does not deviate too far from  RPMA's statistics.

Training was performed on a system equipped with AMD Ryzen Threadripper PRO 3955WX 16C CPU 3.9G running at 64MB cache, and 64GB of CRUCIAL CT8G4DFRA32A RAM clocked at 3200MHz. Program execution was performed solely on the CPU. The CPU-based simulation of the environment, such as moving packets between buffers and calculating the various interference, introduced a CPU-GPU data transfer overhead, negating the advantages of GPU acceleration.

\section{Residual Profit Maximizer Algorithm}\label{GREEDYSEC}
We now present the Residual Profit Maximizer Algorithm (RPMA), whose performance is used as a reference for AARL. The input for RPMA is a traffic matrix for each BS and an interference matrix. It determines which mmWave links will be activated during the next slot, and what power will be used for each activated link.

RPMA maintains a set $L$ of all the links considered for the next slot, and a set $S$ of all the links it has already chosen for the next slot. Initially, $L$ contains all the links and $S$ is empty. During each iteration, RPMA considers one link $l$, which is randomly chosen from $L$. Then, it decides whether to add $l$ into $S$ and using what power.

Adding a link $l$ to $S$ has both positive and negative contributions to the network performance. The positive part, termed \emph{capacity gain} ($C^+$), is due to the fact that packets can be forwarded over this link. The negative part, termed \emph{capacity loss} ($C^-$), is due to the interference between the new link and previously chosen links. The difference between $C^+$ and $C^-$ is referred to as the \textit{residual profit} (RP).

The capacity gain of adding $l$ into $S$ using power $P_\mathrm{R}$ is given by Eq. \ref{eq:effective}, which indicates the number of packets to be transmitted over this link during the next slot if it is activated using the considered power. Then, for each link $l'$ previously added into $S$, RPMA computes the difference between the new and old effective received power of $l'$ while considering the new link $l$ and its power $P_\mathrm{R}$. The capacity loss is the number of packets that could not be routed in the next slot due to the interference of link $l$: $C^- = \sum_{l'  \in \mathrm{S}}{(P_\mathrm{R_\mathrm{eff}}(l'_{old})-P_\mathrm{R_\mathrm{eff}}(l'_{new}))}\frac{C(l')}{P_\mathrm{R}(l')}.$

To summarize, RPMA chooses a random link $l$ from $L$, until $L$ is empty and performs the following:
\begin{enumerate}
    \item For each possible power $p$, it decides what is the RP of adding $l$ with power $p$. 
    \item If the RP is positive for some $p$, $l$ is added to $S$ with $p$ for which the RP is maximum. If there is no $p$ for which the RP is positive, $l$ is not into $S$.
    \item After a decision is made for $l$, it is removed from $L$. 
\end{enumerate} When $L$ is empty, $S$ contains the selected power for the next slot for each mmWave link.

RPMA's time complexity is $O(E^{3}\cdot \ \textit{\#power levels})$, where $E$ is the number of mmWave links in the topology. We implemented RPMA with eleven distinct power levels: between 0 to 1 with 0.1 steps. It's given that a typical dense urban network is expected to have at least a few tens of mmWave links, and that a typical slot length is a few ms \cite{3gpp}. RPMA's time complexity renders it impractical for deployment.

While RPMA is impractical for real-time networks, we use it as follows: Before AARL's training begins, we run RPMA to collect data on the percentage of dropped packets and the average number of slots required to deliver different traffic demand matrices. This information is then used for helping to decide when the training of AARL can be stopped.

\section{Evaluation}\label{EVALSEC}
Since no previous paper addresses the mmWave link scheduling and power control problem, we cannot compare our AARL results to previous work. Still, we compare the results of AARL to those obtained by RPMA, although the time complexity of RPMA renders it impractical for ORAN.

Throughout this section we consider our three example topologies with three different workloads: (a) a uniform workload, where each of the N BSs sends and receives the same amount of data; (b) a few-to-many workload where 10\% of the BSs send data to 90\% of the BSs; and (c) an incast many-to-few workload where 90\% of the BSs send data to 10\% of the BSs. We simulated a mmWave backbone according to the 3GPP use cases \cite{etsi}. The simulation is continuous in the sense that data packets enter the mmWave backbone on a continuous basis, either from the core network or from users via the BSs.

\begin{figure}[t]
\centering
\captionsetup{justification=centering}
\begin{subfigure}{.5\columnwidth}
  \centering
  \captionsetup{justification=centering}
  \includegraphics[width=1\textwidth]{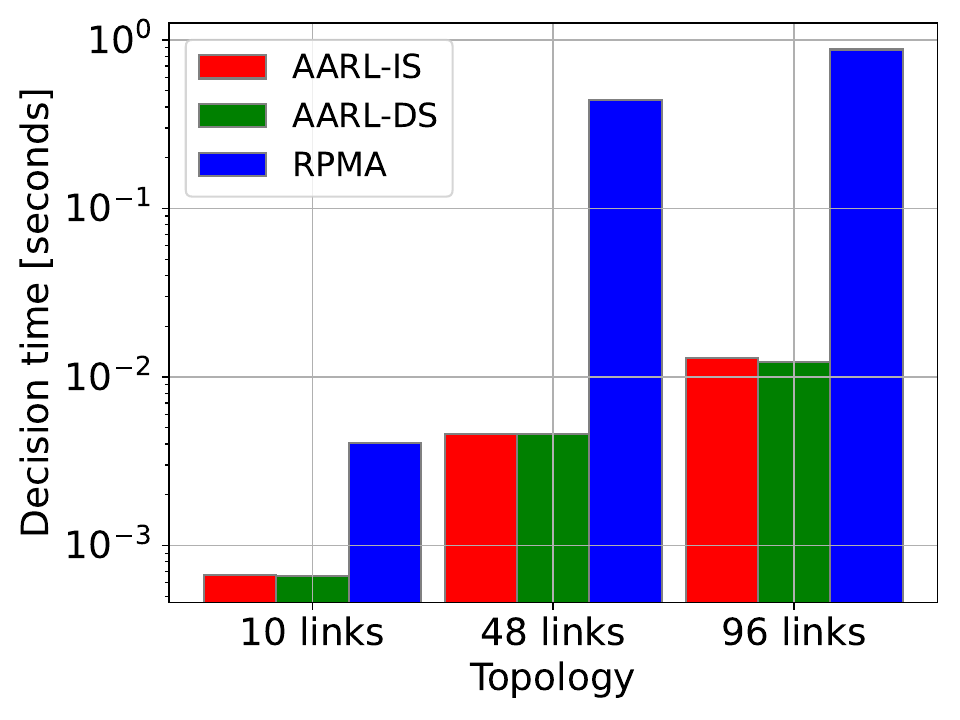}
  \caption{Uniform workload}
  \label{fig:timing-uniform}
\end{subfigure}%
\begin{subfigure}{.5\columnwidth}
  \centering
  \captionsetup{justification=centering}
  \includegraphics[width=1\textwidth]{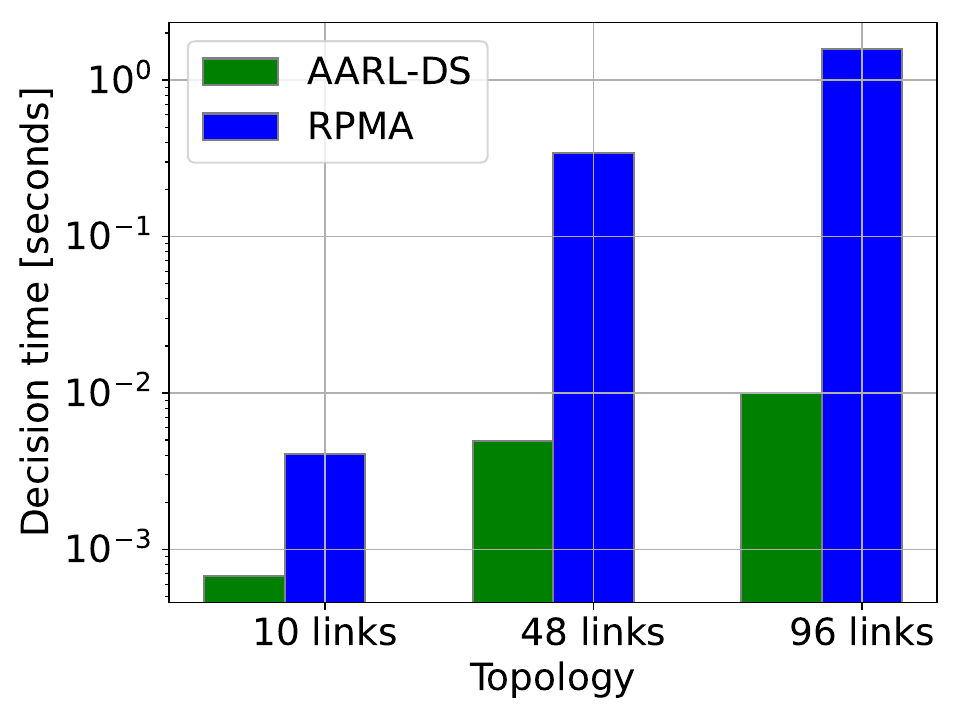}
  \caption{Few-to-many workload}
  \label{fig:timing-90-2-10}
\end{subfigure}
\begin{subfigure}{.5\columnwidth}
  \centering
  \captionsetup{justification=centering}
  \includegraphics[width=1\textwidth]{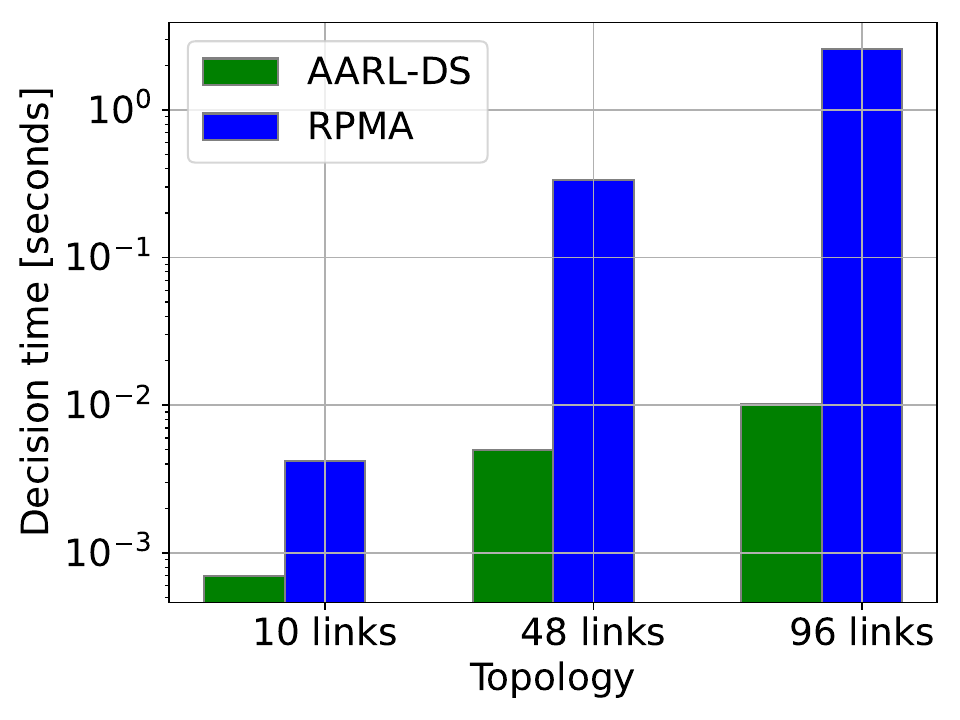}
  \caption{Many-to-few workload}
  \label{fig:timing-10-2-90}
\end{subfigure}
 \caption{The time it takes for AARL and RPMA to make a scheduling decision (logarithmic scale)}
\label{fig:timing-general}
\end{figure}
As explained in Section \ref{MODEL}, the RIC makes scheduling and power control decisions for slot $s+1$ during slot $s$. To this end, the RIC receives a vector demand from each BS via the appropriate interface. The vector demand is then corrected by the RIC, to take into account the links that are activated during slot $s$. The RIC sends its scheduling decision to all BSs via the appropriate interface after deciding which links will be activated during slot $s$.
\begin{table}[t]
\centering
\resizebox{0.65\columnwidth}{!}{
\begin{tabular}{ccccc}
\toprule
\multicolumn{1}{c}{} & \multicolumn{3}{c}{\textbf{Topology}} & \\
\cmidrule(rl){2-5}
\textbf{Workload} & {\makecell{Small\\(10-link)}} & {\makecell{Medium\\(48-link)}} & {\makecell{Large\\(96-link)}}  \\
\midrule
Uniform  & 2,304 & 10,812 & 45,246 \\
Few-to-many  & 1,800 & 3088 & 57,600 \\
Many-to-few  & 1,800 & 13,184 & 57,600 \\ 
\bottomrule
\end{tabular}}
\caption{Evaluation statistics for each combination of topology and workload}
\label{tab:source-destination}
\end{table}
The statistics of the system parameters during the evaluation phase are similar to those used during the training phase. The length of each buffer is set to 650 packets, and the maximum capacity of each mmWave link is 120 packets per slot. For the uniform workload and the 10-link topology, the system starts with 2,304 packets whose sources and destinations are randomly determined using uniform distribution. For the uniform workload and the 48-links topology, the system starts with 10,812 packets whose sources and destinations are uniformly distributed. For the uniform workload and 96-link topology, the system starts with 45,246 packets whose sources and destinations are uniformly distributed. Table \ref{tab:source-destination} summarizes the above numbers, and also presents the number for the other two workloads considered in this section: the few-to-many and the many-to-few. 

The most important property of AARL is its execution time, as shown in Figure \ref{fig:timing-general} on a logarithmic scale. It shows the time it takes for RPMA and AARL to make a scheduling decision for the various topologies and workloads. Despite the fact that AARL employs a larger neural network, when the topology increases, its running time is independent of the topology size. This is because the time AARL needs for making a scheduling decision is the time it takes to compute the feed-forward of the neural-network in the agent's architecture. In contrast, the running time of RPMA strongly depends on the topology size. We can also see that the workload type does not affect the running time of AARL, but it does affect the running time of RPMA.

As discussed in Section \ref{MODEL}, AARL has to meet the time constraints of the slot duration, which is at least 10 ms. It is evident that the running time of RPMA is between 1 and 2 orders of magnitude larger than that of AARL, making it infeasible for real deployment. It is also evident that RPMA is much more affected by an increase of the topology. We can  see that AARL is capable of meeting the time constraint of around 10 ms \cite{3gpp}. Table \ref{tab:timing} summarizes the running times of both algorithms for each topology and workload.

\begin{table}[t]
\centering
\resizebox{0.6\columnwidth}{!}{
\begin{tabular}{cccc}
\toprule
\multicolumn{1}{c}{} & \multicolumn{3}{c}{\textbf{Topology}} \\
\cmidrule(rl){2-4}
\textbf{Workload - Algorithm} & {\makecell{Small\\(10-link)}} & {\makecell{Medium\\(48-link)}} & {\makecell{Large\\(96-link)}}  \\
\midrule
Uniform - RPMA & 4.06 & 438.67 & 876.21 \\
Uniform - AARL-DS  & 0.65 & 4.57 & 12.23 \\
Uniform - AARL-DI  & 0.66 & 4.55 & 12.88 \\
\midrule
Many-to-few - RPMA  & 4.07 & 341.41 & 1584.34 \\ 
Many-to-few - AARL-DS  & 0.67 & 4.96 & 10.01 \\ 
\midrule
Few-to-many - RPMA  & 4.19 & 333.13 & 2588.10 \\ 
Few-to-many - AARL-DS  & 0.69 & 5 & 10.16 \\ 
\bottomrule
\end{tabular}
}
\caption{The time it takes to AARL and RPMA to make a scheduling decision (in ms)}
\label{tab:timing}
\end{table}
The hardware used in the evaluation is the same as used for the training. Since the program is not optimized for GPU execution, as explained in Section \ref{SOLUTION}, we can assume that AARL will make scheduling decisions faster in a real system that will use GPUs, while RPMA will not. This is because matrix calculations performed on GPUs are much faster than on CPUs. 

\begin{figure}[t]
\centering
\captionsetup{justification=centering}
\begin{subfigure}{.50\columnwidth}
  \centering
  \captionsetup{justification=centering}
  \includegraphics[width=1\textwidth]{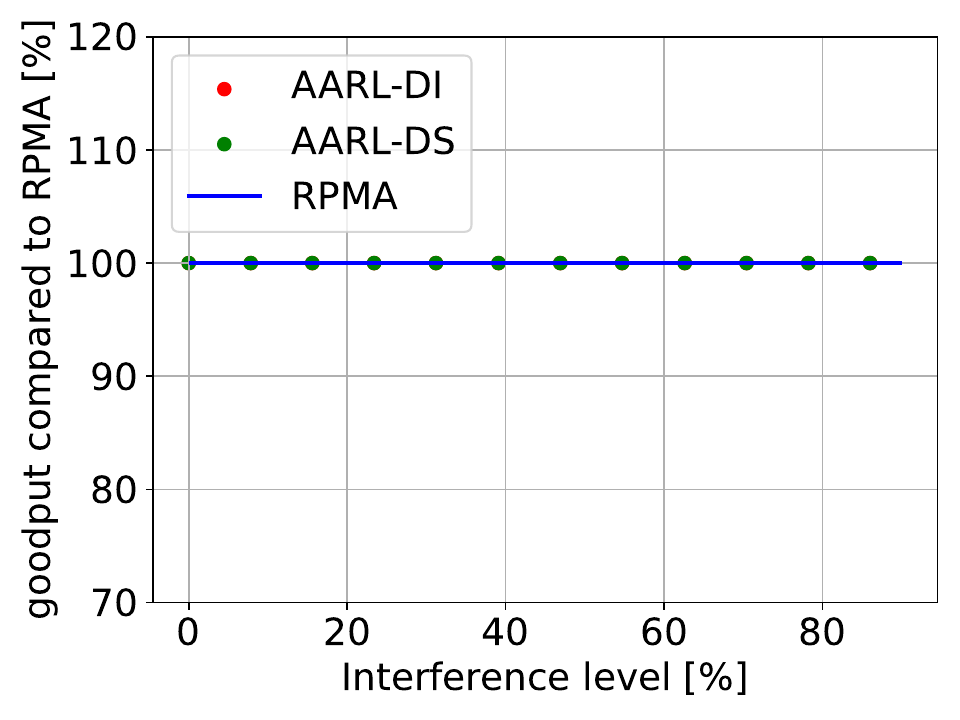}
  \caption{10-link topology}
  \label{fig:10-link-goodput-unifrom}
\end{subfigure}%
\begin{subfigure}{.50\columnwidth}
  \centering
  \captionsetup{justification=centering}
  \includegraphics[width=1\textwidth]{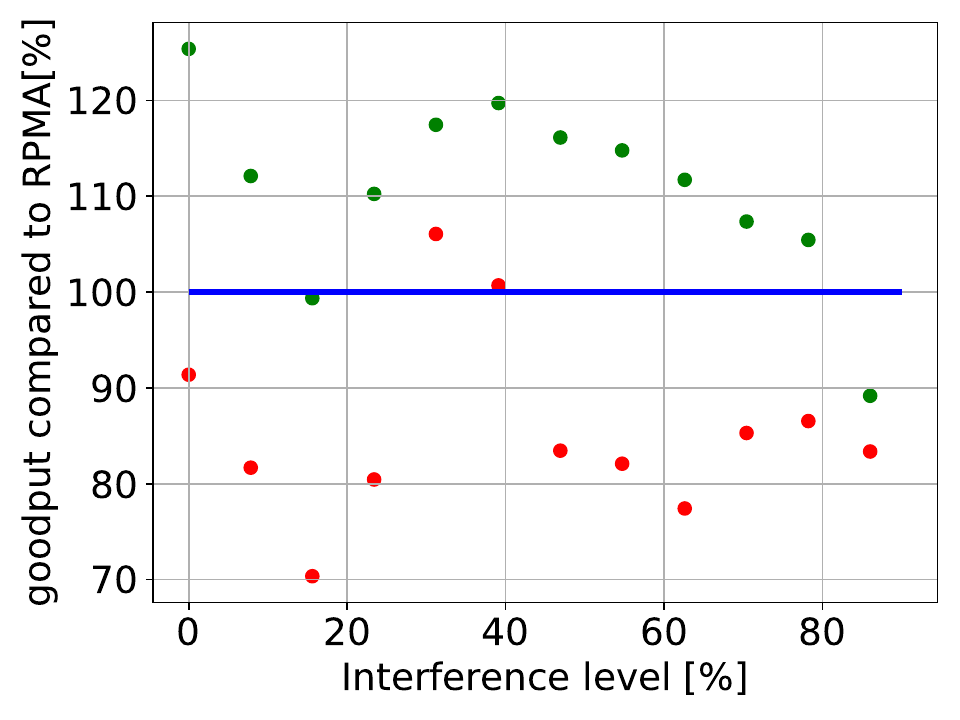}
  \caption{48-link topology}
  \label{fig:48-link-goodput-unifrom}
\end{subfigure}
\begin{subfigure}{.50\columnwidth}
  \centering
  \captionsetup{justification=centering}
  \includegraphics[width=1\textwidth]{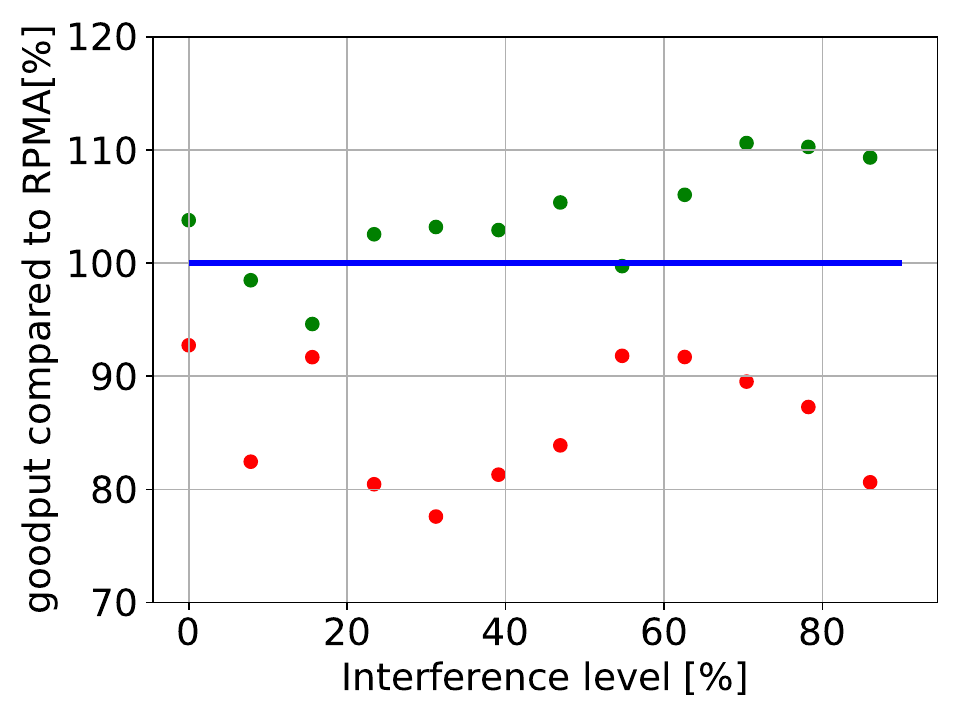}
  \caption{96-link topology}
  \label{fig:96-link-goodput-unifrom}
\end{subfigure}
 \caption{The performance of AARL vs. RPMA for the uniform workload (RPMA/AARL)}
\label{fig:4-gr-uniform}
\end{figure}

Figure \ref{fig:4-gr-uniform} compares the performance of two AARL versions to that of RPMA. These versions are called AARL-DS and AARL-DI. The former uses $\alpha$=10 in its reward function, and the latter uses $\alpha$=1. It shows the goodput obtained by the various algorithms for different interference levels. The interference matrix for each topology is calculated at the beginning of the corresponding scenario. Each time, a different interference level ranging from 0\% to 100\% is used, where 100\% means that only one mmWave link can transmit concurrently. The goodput is defined as 100\% minus the percentage of lost traffic, and it is normalized to the goodput of RPMA (i.e., RPMA's goodput is considered 100\%). Figure \ref{fig:4-gr-uniform}(a) shows that for the 10-link topology the goodput of AARL-DI and AARL-DS is identical to that of RPMA. This is because, the interference introduced by a small-scale topology is minor, hence there is no packet drop for the small number of BSs. For the 48-link and 96-link topologies, Figures \ref{fig:4-gr-uniform}(b) and \ref{fig:4-gr-uniform}(c) show that AARL-DS is better and AARL-DI is worse than RPMA. Unlike RPMA, which only prioritizes moving as many packets considering the interferences, AARL-DS is trained to prioritize the throughput, while also considering the interference between the different mmWave links and the number of dropped packets. Therefore, AARL-DS performs roughly 10\% better then RPMA.

\begin{figure}[t]
\centering
\captionsetup{justification=centering}
\begin{subfigure}{.50\columnwidth}
  \centering
  \captionsetup{justification=centering}
  \includegraphics[width=1\textwidth]{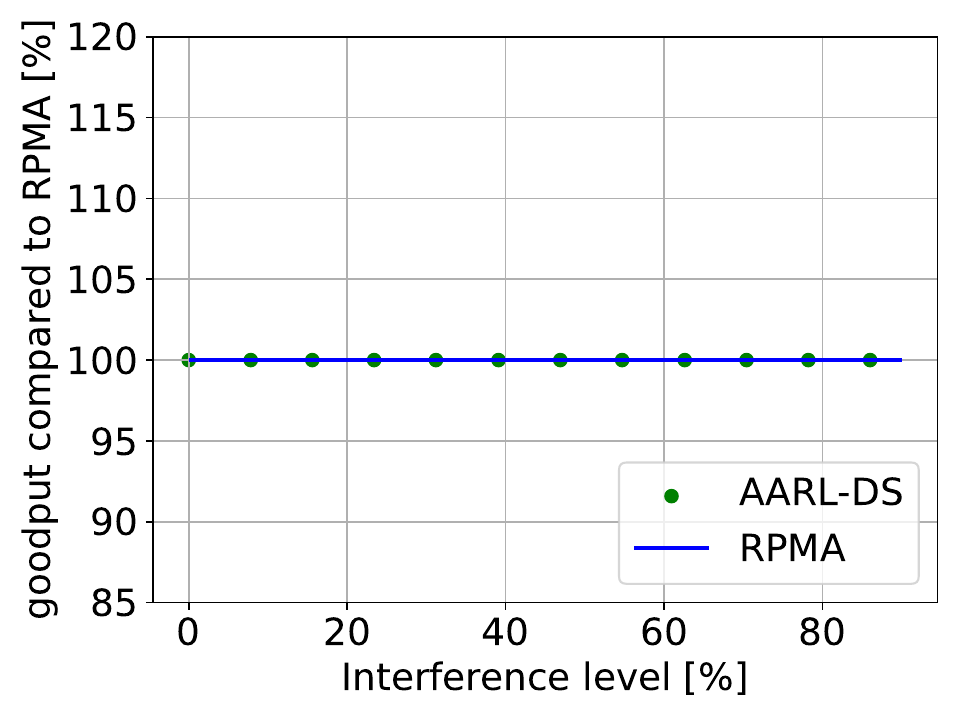}
  \caption{10-link topology}
  \label{fig:10-link-goodput-10-2-90}
\end{subfigure}%
\begin{subfigure}{.50\columnwidth}
  \centering
  \captionsetup{justification=centering}
  \includegraphics[width=1\textwidth]{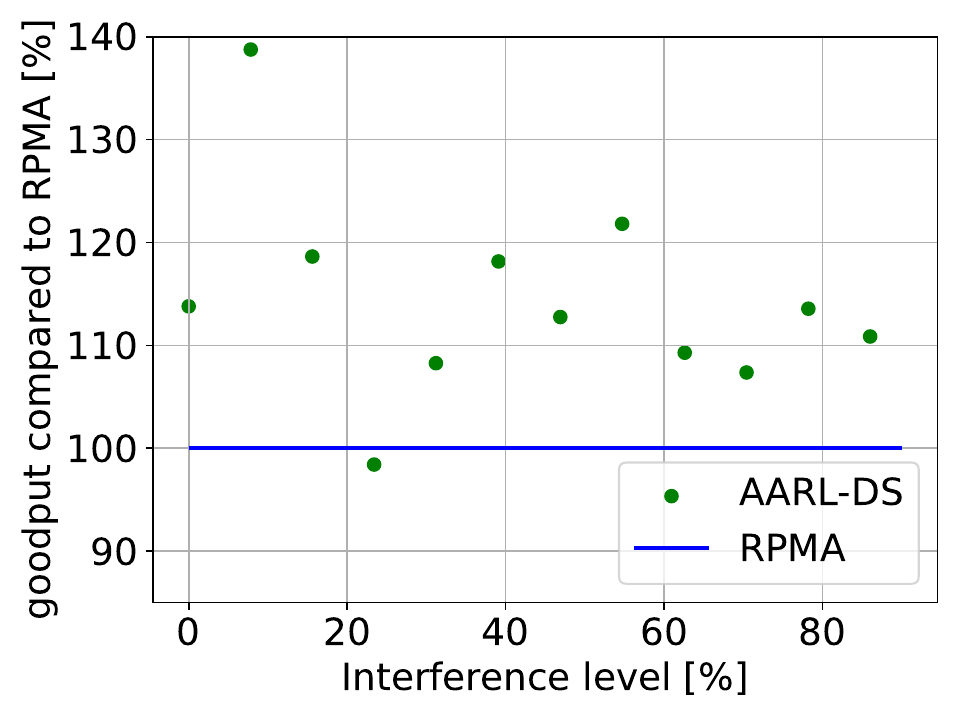}
  \caption{48-link topology}
  \label{fig:48-link-goodput-10-2-90}
\end{subfigure}
\begin{subfigure}{.50\columnwidth}
  \centering
  \captionsetup{justification=centering}
  \includegraphics[width=1\textwidth]{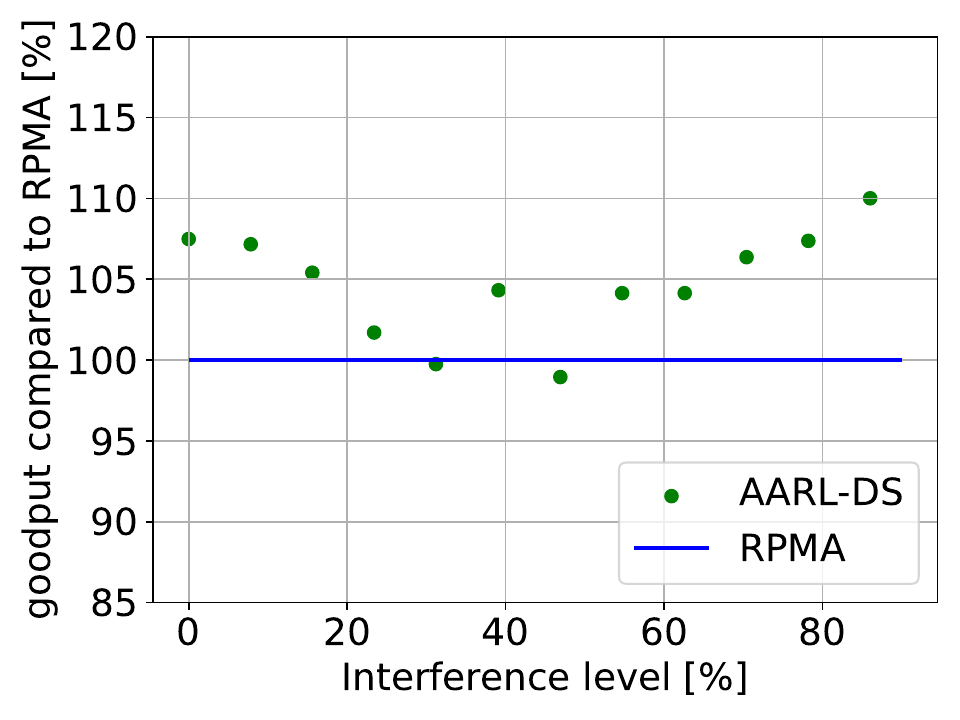}
  \caption{96-link topology}
  \label{fig:96-link-goodput-10-2-90}
\end{subfigure}
 \caption{The performance of AARL vs. RPMA for the few-to-many workload (RPMA/AARL)}
\label{fig:rl_10_to_90-goodput}
\end{figure}

\begin{figure}[t]
\centering
\captionsetup{justification=centering}
\begin{subfigure}{.50\columnwidth}
  \centering
  \captionsetup{justification=centering}
  \includegraphics[width=1\textwidth]{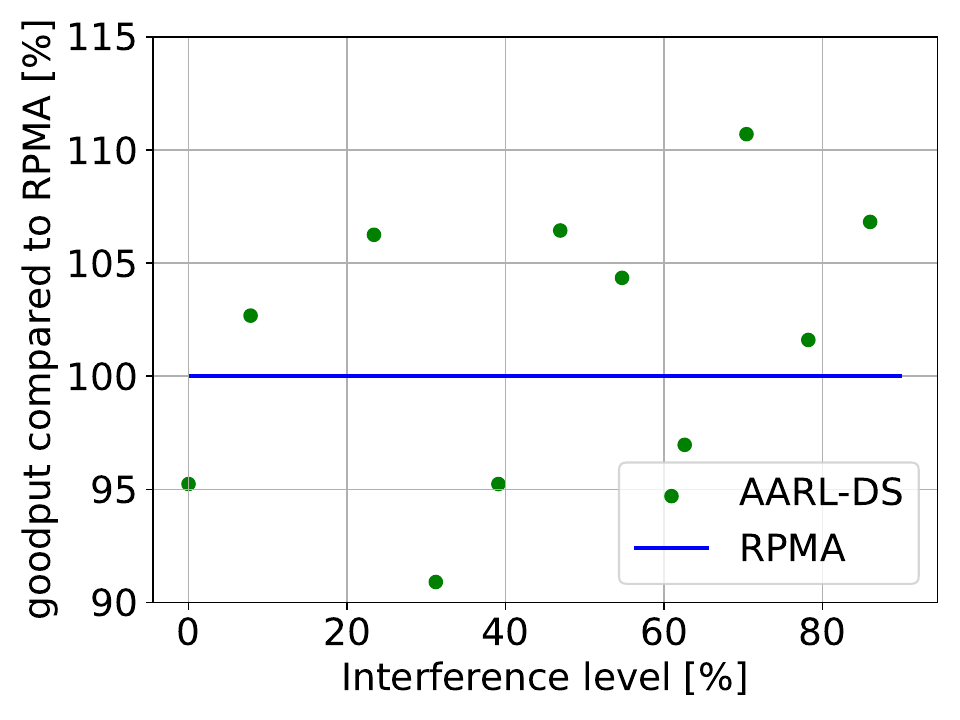}
  \caption{10-link topology}
  \label{fig:10-link-goodput-90-2-10}
\end{subfigure}%
\begin{subfigure}{.50\columnwidth}
  \centering
  \captionsetup{justification=centering}
  \includegraphics[width=1\textwidth]{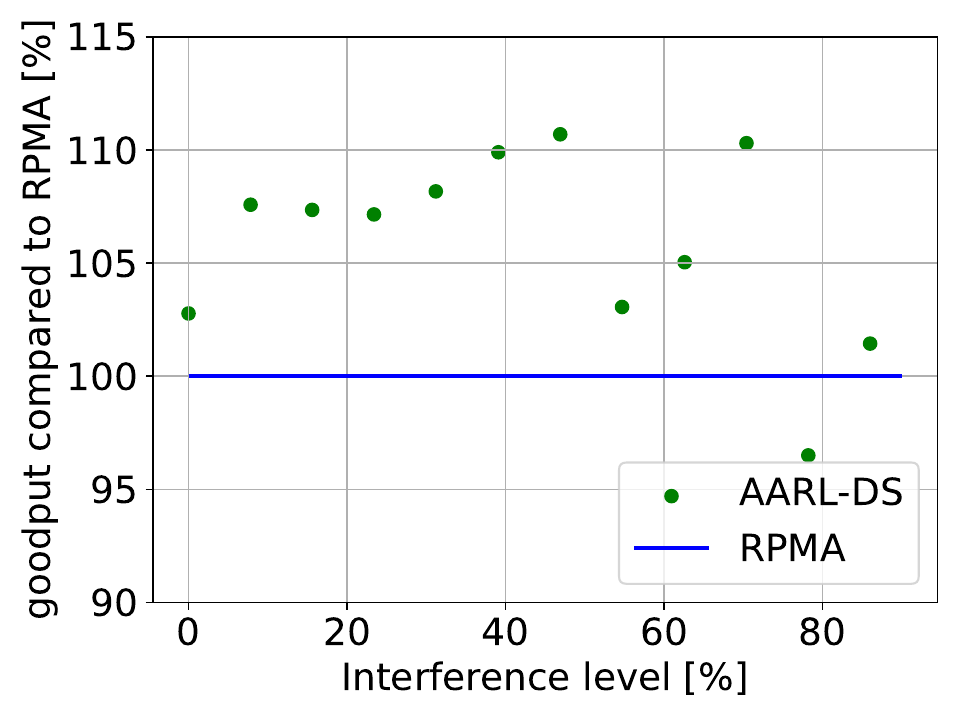}
  \caption{48-link topology}
  \label{fig:48-link-goodput-90-2-10}
\end{subfigure}
\begin{subfigure}{.50\columnwidth}
  \centering
  \captionsetup{justification=centering}
  \includegraphics[width=1\textwidth]{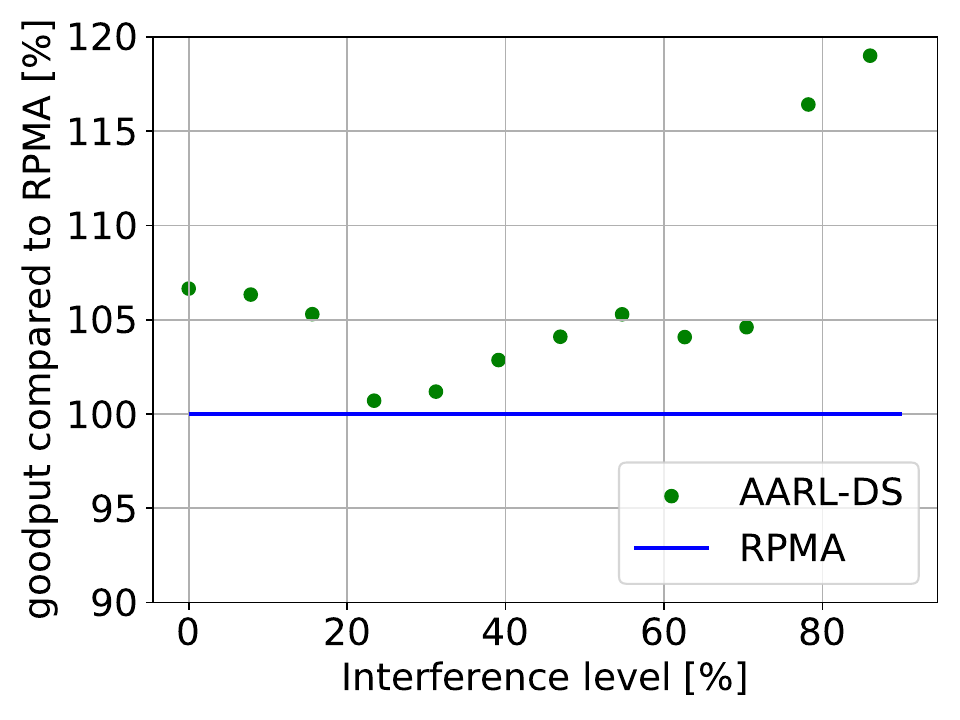}
  \caption{96-link topology}
  \label{fig:96-link-goodput-90-2-10}
\end{subfigure}
 \caption{The performance of AARL vs. RPMA for the many-to-few workload (RPMA/AARL)}
\label{fig:rl_90_to_10-goodput}
\end{figure}

For the additional workloads presented in this section, we consider only AARL-DS and not AARL-DI. Figure \ref{fig:rl_10_to_90-goodput} shows the goodput obtained by AARL-DS and RPMA, for the various interference levels on a workload of few-to-many. The trend is similiar to the uniform workload for all topologies. Figure \ref{fig:rl_90_to_10-goodput} shows the goodput obtained by both algorithms for the  many-to-few workload. Figure \ref{fig:rl_90_to_10-goodput}(a) shows that for the 10-link topology, the goodput of AARL-DS is better than RPMA for the majority of the interference levels. Since it is a small size topology, sending to 10\% of the BSs means that only one BS is a destination. This can result in a high packet loss. However, because AARL-DS is capable of intelligently scheduling mmWave links to maximize throughput while dropping a lower number of packets, it performs better. For the 48-link and 96-link topologies, Figures \ref{fig:rl_90_to_10-goodput}(b) and \ref{fig:rl_90_to_10-goodput}(c) show that the trend is similiar to what we saw in Figures \ref{fig:4-gr-uniform}, \ref{fig:rl_10_to_90-goodput}.

\section{Conclusions}\label{CONCLUDESSEC}
We studied the problem of fast scheduling of real-time flows over a multihop mmWave wireless mesh. We developed a model-free DRL algorithm that does not require any prior knowledge of the topology or the interdependencies between different mmWave links. The proposed algorithm determines which subset of the mmWave links should be activated during each time slot and at what power level, and was demonstrated to work well for multiple workloads and topologies. Its most important property is that it satisfies the timing constraints of 5G mmWave networks without sacrificing performance.

{\footnotesize \bibliographystyle{IEEEtran}
\bibliography{ref}}
\end{document}